# HYAMD High-Resolution Fundus Image Dataset for age related macular degeneration (AMD) Diagnosis

Meisel Meishar[3,4], Cohen Benjamin[1,2], Meital Baskin[3,4], Beatrice Tiosano[3,4] *Behar Joachim A.[1], *Berkowitz Eran[3,4,5]

**Abstract**

The Hillel Yaffe Age Related Macular Degeneration (HYAMD) dataset is a longitudinal collection of 1,560 Digital Fundus Images (DFIs) from 325 patients examined at the Hillel Yaffe Medical Center (Hadera, Israel) between 2021 and 2024. The dataset includes an AMD cohort of 147 patients (aged 54–94) with varying stages of AMD and a control group of 190 diabetic retinopathy (DR) patients (aged 24–92). AMD diagnoses were based on comprehensive clinical ophthalmic evaluations, supported by Optical Coherence Tomography (OCT) and OCT angiography. Non-AMD DFIs were sourced from DR patients without concurrent AMD, diagnosed using macular OCT, fluorescein angiography, and widefield imaging. HYAMD provides gold-standard annotations, ensuring AMD labels were assigned following a full clinical assessment. Images were captured with a DRI OCT Triton (Topcon) camera, offering a 45° field of view and 1960 × 1934 pixel resolution. To the best of our knowledge, HYAMD is the first open-access retinal dataset from an Israeli sample, designed to support AMD identification using machine learning models.

[1] Faculty of Biomedical Engineering, Technion-IIT, Israel. [2]Taub Computer Science Faculty, Technion-IIT, Israel. [3]The Ruth and Bruce Rappaport Faculty of Medicine, Technion-IIT, Israel. [4]Department of Ophthalmology, Hillel Yaffe Medical Center, Israel. [5]The Adelson School of Medicine, Ariel University, Ariel, Israel

## Background

Age related macular degeneration (AMD) is a retinal disease that threatens high-acuity central vision, and a leading cause of blindness. The main risk factor is advancing age, with the severity of vision loss ranging from mild to severe. There is a 25% risk of early AMD and 8% risk of late AMD in patients over the age of 75, with the number of cases expected to increase because of the aging population. The current global prevalence of this disease is about 200 million today and is expected to reach 288 million in 2040 [1]. Due to the high incidence and risk, it is crucial to improve the efficiency of the screening and diagnosis of AMD. In addition, early detection of AMD significantly improves the patient's visual prognosis by allowing timely intervention, which can prevent severe vision loss and enhance the quality of life.

Diagnosis of the disease requires a dilated fundus examination the examiner evaluates the macula for deposits of drusen, pigmentary changes, geographic atrophy and scar formation. Attention is given to the size, number, and distribution of drusen. A complete eye examination is also performed to rule out other coexisting ocular pathologic conditions. The staging of the disease may largely be based on the examination; however, use of a variety of imaging techniques is now considered essential to correlate examination findings and guide management. The etiology of AMD remains incompletely understood, with contributing factors including genetic predisposition, chronic photo-destruction, oxidative stress, and nutritional deficiencies (1, 2). AMD is broadly classified into two categories: Dry AMD (nonexudative) and Wet AMD (neovascular or exudative).

Dry AMD is characterized by the accumulation of drusen and progressive atrophy of the retinal pigment epithelium (RPE), leading to gradual vision loss (3). In advanced stages, large areas of atrophy and extensive drusen deposits are visible under ophthalmoscopy. Wet AMD, on the other hand, is marked by pathological neovascularization beneath the RPE, resulting in exudation, hemorrhage, and scarring. This form progresses rapidly and causes irreversible damage to photoreceptors, leading to severe and rapid vision loss if untreated (4)."

The Age-Related Eye Disease Study (AREDS) developed a classification system to assess the severity of AMD, categorizing it into mild, moderate, and severe stages based on clinical features. Mild AMD is characterized by the presence of small drusen (<63 μm) without significant visual symptoms. Moderate AMD involves medium-sized drusen (≥63 μm and <125 μm) or pigmentary abnormalities, with a higher risk of progression. Severe AMD is marked by large drusen (≥125 μm), geographic atrophy, or neovascular AMD, often accompanied by significant visual impairment. This classification aids in predicting disease progression and guiding management strategies (5)

Recent advancements in deep learning (DL) have revolutionized the field of automated age-related macular degeneration (AMD) detection, with digital fundus imaging (DFI) serving as a primary modality for these applications (6). However, a persistent limitation in the existing body of research is the reliance on reference labels derived solely from DFI evaluations, rather than comprehensive ophthalmic examinations (7). This approach reduces the detection task to a subjective interpretation of retinal appearance, which is inherently prone to variability and error. Conditions such as diabetic retinopathy and central retinal vein occlusion may mimic AMD features, further complicating the identification process. Consequently, DL models trained exclusively on DFI data are susceptible to inheriting biases from inconsistent annotations, subjective judgments, and unverified cases, potentially misrepresenting true clinical manifestations of AMD.

To address these critical issues, the Hillel Yaffe AMD Dataset (HYAMD) was specifically developed to provide a robust foundation for AMD detection and classification tasks. Unlike most existing datasets, HYAMD employs gold-standard annotations based on comprehensive ophthalmic evaluations, including optical coherence tomography (OCT), rather than relying solely on subjective DFI-based labeling. By incorporating a comprehensive ophthalmic examination together with multimodal diagnostic data, HYAMD aims to enhance the reliability of DL models and mitigate biases, ensuring that algorithmic outputs align more closely with clinical realities. This resource represents a pivotal step toward improving the accuracy and generalizability of AI-driven AMD detection systems, ultimately benefiting both researchers and clinicians. HYAMD is also, to the best of our knowledge, the first open access retinal dataset from an Israeli population sample.

## Methods

This study was approved by the Helsinki Committee at the Hillel Yaffe Medical Center (Helsinki approval number: 0029-24-HYMC). All identifiable patient information was removed to ensure patient privacy.

## Study Cohort

The dataset was curated by the Hillel Yaffe Ophthalmology Department Retina Unit, Hadera, Israel. DFIs of both eyes of each patient were captured using a TOPCON DRI OCT Triton retinal camera, with a 45° FOV. The AMD cohort includes 188 patients (95 males, 93 females) aged 57–95, while the control group comprises 192 diabetic retinopathy (DR) patients (113 males, 79 females) aged 24–92 (Figure 1).

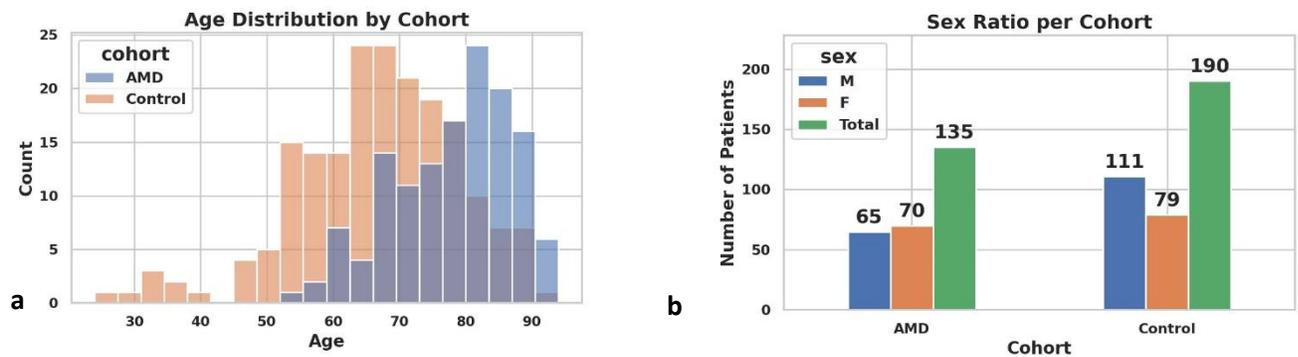

Figure 1 – Patient distribution by age (a) and sex (b)

## Data preparation

All DFIs were deidentified, ensuring that any personal identifiers were removed. To maintain consistency, images were cropped to a square format by removing black borders.

## Labeling

HYAMD employs gold-standard annotations, meaning that AMD labels were assigned based on a full ophthalmic examination, rather than being inferred solely from DFIs. Patients are diagnosed with AMD based on a comprehensive ophthalmic examination, which includes visual acuity (VA) assessment, intraocular pressure (IOP) measurement, anterior and posterior segment evaluation and posterior pole assessment using a 78 diopter lens while the pupil is dilated. These examinations are supported by OCT scans, including OCT angiography-. The AMD cohort is divided in two subgroups: early AMD, with a label AMD=1, and intermediate-to-late AMD, with a label of AMD=2. The intermediate-to-late group either presents with neovascular AMD (NVAMD), visible geographic atrophy (GA), or with the presence of large drusen (LD) (Figure 2). Furthermore, all patients in this cohort were followed up in Retina clinic for at least a year validating the labeling accuracy. Each patient had at least three visits a year, during which they underwent clinical examinations, OCT, and fundus photography as part of their routine follow-up. If the stage of an eye changed during the follow-up period, all subsequent images were labeled with the new AMD stage. Patients with erroneous diagnoses were removed from the dataset. Non-AMD DFIs were collected from patients examined in the Retina clinic for DR who did not have a concurrent diagnosis of AMD or other eye diseases. The diagnosis of DR was based on clinical examination, macular OCT, and, where relevant, fluorescein angiography and widefield imaging.

All examinations were carried out by professional Retina specialists, while the images were taken by trained and experienced technicians.

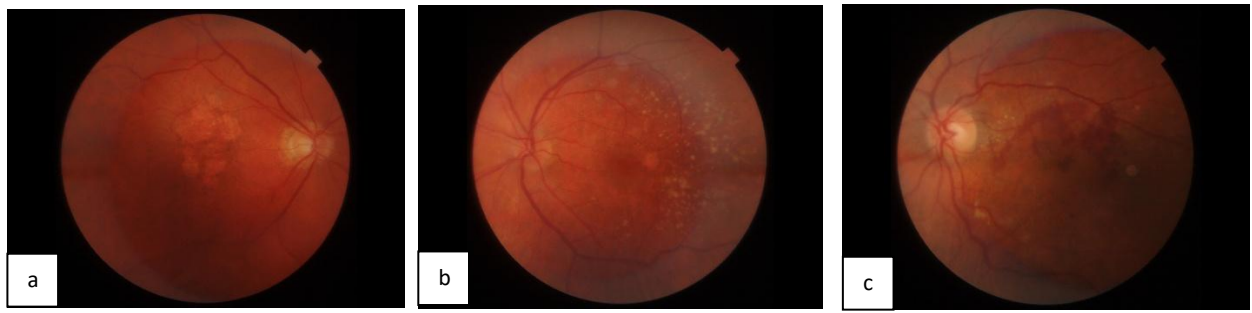

Figure 2 – Example fundus images of GA (a), LD (b) and nvAMD (c).

**Data Description**

The structure of the dataset is as follows:

**Images**

The dataset consists of DFIs, including:

☐ AMD DFIs

☐ non-AMD DFIs

All DFIs are stored in JPG format with a 1:1 aspect ratio. The naming convention follows the format x_y_z.jpg, where:

☐ x represents the patient ID.

☐ y represents the date the image was taken.

☐ z represents the laterality.

For example, 981616684_11-09-2023_R

**Labels**

The labels file contains the following columns:

| Column Name | Description |
| --- | --- |
| **Image ID** | The filename of the DFI. |
| **Patient ID** | The patient ID. |
| **Eye side** | A binary classification: OD ( right eye) or OS (left eye) |
| **AMD** | A nulticlass classification: 0 – non AMD (DR control), 1 – early AMD, 2- Intermediate to late stage AMD |
| **Sex** | A binary classification: Male (M) or Female (F) |
| **Age** | In Years |

**Usage Notes**

**Usage**

This dataset can serve as a benchmark dataset for evaluating existing AMD diagnosis models or as a resource for training new deep learning models for automated AMD detection. In addition this dataset can bes a valuable resource for validating AI models across diverse demographics The dataset's dual focus on AMD and DR opens additional doors for comparative analysis, possibly used to develop hybrid diagnostic models

## Limitations

The dataset has several limitations:

Patients' medical history beyond AMD diagnosis is not provided.

The dataset has limited demographic diversity, as it was collected using a single camera model with a fixed 45° FOV and includes data from a single ethnic group.

## Release Notes

Version 1.0.0: Initial release.

## Ethics

The authors declare no ethics concerns. This project was approved by the Helsinki Committee at the Hillel Yaffe medical center (Helsinki approval number 0029-24-HYMC).

## Conflicts of Interest

The authors have no conflicts of interest to declare.